\documentclass[sn-mathphys-num]{sn-jnl}
\usepackage{graphicx}
\usepackage{multirow}
\usepackage{amsmath,amssymb,amsfonts}
\usepackage{amsthm}
\usepackage{mathrsfs}
\usepackage[title]{appendix}
\usepackage{xcolor}
\usepackage{textcomp}
\usepackage{manyfoot}
\usepackage{booktabs}
\usepackage{algorithm}
\usepackage{algorithmicx}
\usepackage{algpseudocode}
\usepackage{listings}

\usepackage{soul,color}
\usepackage[multiple]{footmisc}
\usepackage{xurl}
\usepackage{mathtools}
\usepackage{caption}
\usepackage{subcaption}

\theoremstyle{thmstyleone}

\theoremstyle{thmstyletwo}

\theoremstyle{thmstylethree}

\begin{document}

\title[Right to be Forgotten in the Era of Large Language Models: Implications, Challenges, and Solutions]{Right to be Forgotten in the Era of Large Language Models: Implications, Challenges, and Solutions}

\author*[1,2]{\fnm{Dawen} \sur{Zhang}}\email{Dawen.Zhang@data61.csiro.au}

\author[1]{\fnm{Pamela} \sur{Finckenberg-Broman}}

\author[1]{\fnm{Thong} \sur{Hoang}}

\author[1,2]{\fnm{Shidong} \sur{Pan}}

\author[1,2]{\fnm{Zhenchang} \sur{Xing}}

\author[1]{\fnm{Mark} \sur{Staples}}

\author[1]{\fnm{Xiwei} \sur{Xu}}

\affil*[1]{\orgname{CSIRO's Data61}, \orgaddress{\country{Australia}}}

\affil[2]{\orgdiv{School of Computing}, \orgname{Australian National University}, \orgaddress{\country{Australia}}}


\abstract{The Right to be Forgotten (RTBF) was first established as the result of the ruling of Google Spain SL, Google Inc.\ v AEPD, Mario Costeja González, and was later included as the Right to Erasure under the General Data Protection Regulation (GDPR) of European Union to allow individuals the right to request personal data be deleted by organizations. Specifically for search engines, individuals can send requests to organizations to exclude their information from the query results. It was a significant emergent right as the result of the evolution of technology. With the recent development of Large Language Models (LLMs) and their use in chatbots, LLM-enabled software systems have become popular. But they are not excluded from the RTBF. Compared with the indexing approach used by search engines, LLMs store, and process information in a completely different way. This poses new challenges for compliance with the RTBF. In this paper, we explore these challenges and provide our insights on how to implement technical solutions for the RTBF, including the use of differential privacy, machine unlearning, model editing, and guardrails. With the rapid advancement of AI and the increasing need of regulating this powerful technology, learning from the case of RTBF can provide valuable lessons for technical practitioners, legal experts, organizations, and authorities.}

\keywords{GDPR, Right to be forgotten, Privacy, Large Language Models}

\maketitle

\section{The Legal Principles behind Right to be Forgotten}\label{sec1}

Privacy is a fundamental human right. This is affirmed by a collection of documents, including the International Covenant on Civil and Political Rights \mbox{(ICCPR)} Article 17\footnote{\url{https://www.ohchr.org/en/instruments-mechanisms/instruments/international-covenant-civil-and-political-rights\#article-17}}, Universal Declaration of Human Rights (UDHR) Article 12\footnote{\url{https://www.un.org/en/about-us/universal-declaration-of-human-rights}}, and the European Convention on Human Rights (ECHR) Article 8\footnote{\url{https://fra.europa.eu/en/law-reference/european-convention-human-rights-article-8-0}}. It is explicitly declared in ECHR that "Everyone has the right to respect for his or her private and family life, home, and communications".

\textit{``Everyone has the right to the protection of personal data concerning him or her. Such data
must be processed fairly for specified purposes and on the basis of the consent of the person
concerned or some other legitimate basis laid down by law. Everyone has the right of access
to data which has been collected concerning him or her, and the right to have it rectified.
Compliance with these rules shall be subject to control by an independent authority.''}
\\
--- Article 8 - Protection of personal data, EU Charter of Fundamental Rights.

These Fundamental rights are echoed in secondary laws, such as the General Data Protection Regulation (GDPR) and Directive 95/46 (Data Protection Directive), which, according to Articles 1-3 of GDPR, protect the fundamental rights and freedoms of natural persons, their right to privacy with respect to the processing of personal data, and of removing obstacles to the free flow of such data.

Most nations protect privacy on different levels. In some, the Right to Privacy is protected through constitutional provisions; in others, by national laws, and entering international treaties. Notably, the extent of protection afforded to individuals, the types of data that can be accessed and utilised, and the parties covered by data protection laws differ significantly across jurisdictions. This includes determining whose data can be accessed, processed, and exploited within legal bounds and whether data protection encompasses both private and public entities engaging in exploitation and surveillance activities.

\subsection{The Rights of the Data Subject}

As a result of EU Data Protection Reform\footnote{\url{https://commission.europa.eu/law/law-topic/data-protection/reform\_en}}, several sub-rights were included under Chapter 3 the Rights of the Data Subject in \textit{General Data Protection Regulation} (GDPR)\footnote{The European Parliament and of the Council Regulation (EU) 2016/679 of 27 April 2016 on the protection of natural persons with regard to the processing of personal data and on the free movement of such data, and repealing Directive 95/46/EC (General Data Protection Regulation): \url{https://eur-lex.europa.eu/legal-content/EN/TXT/?uri=CELEX:32016R0679}} of European Union. It furthers the protection of these rights by codifying a collection of sub-rights (Art.~12--23) into the law, notably the Right to be Informed (Art.~13, Art.~14), Right of Access (Art.~15), Right to Rectification (Art.~16), and Right to Erasure (Art.~17).

\begin{itemize}
    \item \textit{Right to be Informed (Art.~13, Art.~14)}
    Individuals have the right to be informed when their data is collected and used. If the data is obtained from the data subject directly, the data subject must be informed at the time when personal data are obtained; and if the data is obtained from other sources, the data subject must be informed within a reasonable period of time which should not exceed one month.
    \item \textit{Right of Access (Art.~15)}
    Individuals have the right to request information about the processing of their personal data. Specifically, the information may include whether or not their personal data is processed, access to the personal data, and information about the processing of personal data such as purpose, duration, etc. The right of access also allows the data subjects to learn about their personal data being processed before practicing other rights where the access to their personal data being processed is necessary or helpful.
    \item \textit{Right to Rectification (Art.~16)}
    The data subjects have the right to have their inaccurate personal data rectified from the controller.
    \item \textit{Right to Erasure (Art.~17)}
    The data subjects have the right to request erasure of their personal data from the controller without undue delay. The right may also apply to circumstances where the personal data is inaccurate\footnote{TU and RE v Google LLC: \url{https://eur-lex.europa.eu/legal-content/en/TXT/?uri=CELEX:62020CJ0460}}. The undue delay is commonly regarded as one month, and may be extended by 2 months for complex or multiple requests where the data subjects shall be informed\footnote{Responding to requests - Data protection under GDPR: \url{https://europa.eu/youreurope/business/dealing-with-customers/data-protection/data-protection-gdpr/index_en.htm}}.
    
\end{itemize}

These rights represents a progressive journey towards greater personal autonomy over one's data. It starts with being informed about data collection practices, proceeds through accessing and verifying the data, enables corrections to be made, and ultimately, under certain conditions, allows for the erasure of data.

The Right to be Informed, Right of Access, and Right to Rectification have their origins in the Data Protection Directive (Directive 95/46/EC), established by the European Parliament and Council on October 24, 1995. These rights were specifically outlined across various sections of the Directive, including recital 25 and Articles 6, 7, 10, 11, 12, and 14. The Right to Erasure, which is also called the Right to be Forgotten (RTBF) was established in the case decision of \textit{Google Spain SL, Google Inc. v Agencia Española de Protección de Datos (AEPD)
and Mario Costeja González (2014)}~\cite{2014rtbfcase}, and was later codified in GDPR.

The Right to be Forgotten is widely considered a crucial evolution in privacy rights in the digital age. It addresses the unique challenges posed by the internet and digital technologies, by offering a form of control for individuals over their digital footprint. It is considered an emergent right~\cite{werro2020right} in response to the rise of internet technologies such as search engines. This exemplifies the evolution of privacy rights, indicating the need to adapt legal and ethical frameworks to address the challenges posed by the new technologies.

Given the rapid advancement of AI and the privacy concerns it raises, it becomes crucial to examine the Right to be Forgotten. By employing the comparative approach to analyse the impact of both the internet technologies and AI on privacy we aim to gain insight into how privacy interacts with new technologies, and how it will shape the evolving landscape of AI technologies.

\subsection{The Legal Principles behind Right to be Forgotten}

The \textit{Right to be Forgotten} (RTBF) is a critical aspect of the fundamental human right to Privacy. RTBF evolved and established itself as a legal principle from the case law of the Court of Justice of the European Union (the CJEU) within the European Union (EU) in response to the rise of the data-driven society, where vast quantities of data and information are gathered, processed, stored, and exchanged for diverse motives.

The \textit{Right to be Forgotten} (RTBF) evolved and was established from the case of \textit{Google Spain SL, Google Inc. v Agencia Española de Protección de Datos (AEPD)
and Mario Costeja González (2014)}~\cite{2014rtbfcase}. An advertisement related to the data subject was published in a newspaper and indexed by Google, which the data subject later discovered by searching his name. The data subject submitted a request to remove his name from the newspaper and Google search engine; however, both authorities rejected the request. A legal case between the data subject and Google went to the Court of Justice of the European Union. Their ruling required the search engine to remove the personal data of individuals upon their requests. The ruling referred to multiple principles, notably privacy, legitimate interests, and balancing of interests.

\begin{itemize}
    \item \textbf{Privacy:} The ruling cites Articles 7 and 8 of the Charter of Fundamental Rights of the European Union (the Charter), declaring that the processing of personal data should respect the privacy of data subjects. Interestingly, the ruling explicitly mentioned that the personal information of the data subject would not be ubiquitously available and interconnected without the existence of the internet and specifically search engines in modern society.

    \item \textbf{Legitimate interests:} The ruling found that the operators of search engines are the controllers of their data, as their processing of data has different legitimate interests and consequences from the original publishers of information.

    \item \textbf{Balancing of interests:} The ruling acknowledges that the processing of personal information is necessary if the processing is for legitimate interests, but that such interests can be overridden by the interests of the data subject's fundamental rights and freedoms.

\end{itemize}

Moreover, RTBF is not an absolute right. GDPR explicitly defined six grounds for the erasure of personal data, including situations where data collection or processing is no longer relevant or consent has been revoked. Additionally, there are four circumstances in which this right shall not apply, such as when it conflicts with freedom of expression or certain public interests.
Other jurisdictions also recognize such rights, including Argentina, India, and South Korea.

Since the establishment of the case, Google published an online form\footnote{\url{https://reportcontent.google.com/forms/rtbf}} for users to request results related to them being delisted.
According to their published paper~\cite{5yrtbf}, they processed 3,231,694 requests about RTBF during the five years from May 30th, 2014, to May 31st, 2019. They found that news, government, social media, and directory sites were the most targeted for submitting RTBF requests.

\section{Data Practices of Advanced AI Technologies}
\label{legal-requirements-data-practices}

The success in advanced AI technologies can be attributed to the extensive datasets on which these models are trained. As discussed previously, recent advancements in Generative AI, such as Large Language Models, possess the ability to reproduce their training data. This capability diverges significantly from the prediction or regression tasks performed by legacy AI models. Such developments raise privacy concerns, particularly regarding the rights of the data subjects whose information may be output by these generative technologies.

\subsection{The Training of Large Language Models}

Large Language Models (LLMs) are language models based on deep neural networks (DNNs) with billions of parameters trained on vast amounts of data~\cite{zhao2023survey}. The training data for these LLMs consists of a large proportion of public data on the internet. For example, Common Crawl data is 60\% of the training data used in GPT-3~\cite{brown2020language}; 50\% of PaLM's training dataset is social media conversations~\cite{palm}; and OpenAI and Google have extensively used Reddit user posts in their large language models~\cite{radford2019language,yang2018learning,gpt4techreport}. Moreover, these AI companies such as OpenAI may also collect user interactions with large language models for future training\footnote{New ways to manage your data in ChatGPT: \url{https://openai.com/blog/new-ways-to-manage-your-data-in-chatgpt}}.

Unlike previous pre-trained language models, i.e., ELMo~\cite{edunov2019pre} and BERT~\cite{wang2019bert}, recent LLMs have become more end-user-facing, which requires users to interact with them through a prompting interface. Specifically, users may need to understand how LLMs work and guide LLMs to understand their problems. For example, GPT-3~\cite{floridi2020gpt} popularized the paradigm of prompting, in which users input a carefully crafted text prompt, and by completing the text, GPT-3 may be able to output useful information for certain tasks. Some researchers called such capabilities that only exist in large models but not in small models the "emergent abilities" of LLMs~\cite{wei2022emergent}. To exploit such capabilities of LLMs to follow human instructions, more advanced models, such as InstructGPT and ChatGPT, rely on human feedback during training, which involves a reward function trained on manually generated prompt-response pairs through \textit{Reinforcement Learning from Human Feedback} (RLHF), to steer the LLM to follow human instructions~\cite{rlhf}. Based on these techniques, technology companies and the community have launched chatbots based on LLMs, including OpenAI's ChatGPT\footnote{\url{https://openai.com/blog/chatgpt}}, Google's Gemini\footnote{\url{https://blog.google/technology/ai/google-gemini-ai/}}, Meta's LLaMA\footnote{\url{https://ai.facebook.com/blog/large-language-model-llama-meta-ai/}}, and Anthropic's Claude\footnote{\url{https://www.anthropic.com/product}}. LLMs have been embedded into other tools, including search engines like Microsoft's Bing\footnote{\url{https://www.bing.com/new}} and GitHub's Copilot\footnote{\url{https://github.com/features/copilot}}. Moreover, some LLMs have been given access to other tools, such as Google's Gemini\footnote{\url{https://blog.google/technology/ai/google-gemini-ai/}} or OpenAI's ChatGPT Plugins\footnote{\url{https://openai.com/blog/chatgpt-plugins}}.

\subsection{Issues of LLMs related to Personal Data}
\label{legal-requirements-llm-issues}

However, though LLMs are impressive and considered a disruptive technology by many, they suffer major issues related to data. Training data memorization and hallucination are two typical issues.

\noindent\textbf{Training Data Memorization.} It has been observed that LLMs may memorize personal data, and this data can appear in their output~\cite{gpt2extract, carlini2023extracting}. We confirm that such behaviours of LLMs can be easily reproduced, and memorization may happen without explicitly asking for related information (e.g.\ ``Please tell me about [person name]''). Even though companies try to remove personal data from training datasets, there could still be personal data contained in the training dataset\footnote{\url{https://openai.com/blog/our-approach-to-ai-safety}}.

\noindent\textbf{Hallucination.} Large language models may output factually incorrect content, which is known as "hallucination"~\cite{hallucination}, and producing such output does not require factually incorrect information to be present in the training dataset. Even if the context is given, the LLM-based generative search engines are possible to give incorrect citations, or draw wrong conclusions from the context~\cite{searchenginehallucination}. Even though companies adopt guardrails and sophisticated prompts to minimize the rate of hallucination, hallucination is still unavoidable.

We produced an example of hallucination using ChatGPT (GPT-4) on 1 April 2024 as shown in the links\footnote{\url{https://chat.openai.com/share/104fabb7-189a-4c74-96c5-93924fb96d00}}\footnote{\url{https://chat.openai.com/share/7cc08e6f-aa71-47b3-8c29-cb5b607725ab}}. With identical prompts as input asking about a late professor in medicinal chemistry, ChatGPT provided different but both inaccurate information in two conversations, including the authorship information and death date.

\section{Large Language Models and the Rights of the Data Subject}
One of the most significant right within the Rights of the Data Subject of the GDPR is the Right to be Forgotten (RTBF), which was an emergent right in the digital era established by a case related to search engines. To understand the implications of advanced AI technologies on privacy, we make a comparison between LLMs and search engines regarding the Right to be Forgotten.

\subsection{Comparing LLMs and Search Engines}
\label{sec:comparison}
The RTBF was first established via a case targeting search engines. Interestingly, many on social media debate whether ChatGPT will replace search engines\footnote{\url{https://www.nytimes.com/2022/12/21/technology/ai-chatgpt-google-search.html}}. We make a technical comparison between LLMs and search engines and discuss whether the operations of LLMs align with or diverge from those of search engines in contexts relevant to RTBF. We identify three major similarities and three main differences between LLMs and search engines mentioned as follows.

\paragraph{Similarities}
\begin{itemize}
    \item \textbf{Organizing web data.}
    LLMs and search engines have sourced data from the web. Specifically, LLMs, as deep neural networks built upon layers of weights, are trained on datasets that include a substantial proportion of data scraped from the web. On the other hand, search engines use crawlers to systematically scrape web pages and index the content, enabling their ability to retrieve and present relevant information in response to user queries.
    \item \textbf{Used to access information.}
    Users often employ LLMs and search engines to access information. While LLMs are trained on a vast amount of online data and output responses based on patterns learned from this information in a generative way, search engines search through online information via indexes and present the responses as direct links to the source materials.
    \item \textbf{Intertwined with each other.} LLMs have been embedded into search engines, while search engines are also now embedded into LLMs. Microsoft has incorporated GPT-4 into its Bing search engine as Copilot, enhancing its user experience with more contextually relevant information in natural language. Google has embedded search engine capabilities into its LLM, Gemini, allowing the model to access real-time information from the web to inform its responses.
\end{itemize}

\paragraph{Disimilarities}
\begin{itemize}
    \item \textbf{Predicting words vs. indexing information.}
    LLMs are trained to predict the next word in a sentence, and the relationship between words does not necessarily reflect the actual information of the trained models. Search engines are created to collect, index, and rank relevant web pages based on user queries. 
    \item \textbf{Conversational chatbots vs. search box.} One trend of LLMs is their deployment within conversational chatbot interfaces, designed to engage users in multi-round dialogues. In contrast, search engines provide service through a more traditional interaction model, relying on a search box interface where users input queries. The search engine processes these queries to return a list of web pages ranked by relevance.

\end{itemize}

Overall, LLMs have similar source of data to search engines, and the datasets used to develop these models may contain personal data, causing privacy concerns. ChatGPT, a chatbot built upon GPT, has become the fastest-growing application in history\footnote{\url{https://www.reuters.com/technology/chatgpt-sets-record-fastest-growing-user-base-analyst-note-2023-02-01/}}, which may also place LLMs in a similar position as search engines mentioned in the court ruling of RTBF that they may make the information more ubiquitously available and interconnected. Similar to search engines, LLMs do not share the same legitimate interests as the original publishers from whom the data were crawled. Instead, these data are used by organizations to train models and provide chatbot services to users. The use of these data can also lead to varied consequences for the original publishers. Therefore, organizations training LLMs should be considered data controllers.

The balancing of interests allows the use of data to be justified by legitimate interests such as research purposes or freedom of speech. While OpenAI labels ChatGPT as a ``low key research preview''~\footnote{\url{https://www.nytimes.com/2023/12/05/technology/ai-chatgpt-google-meta.html}} and Google marks Bard as an ``experiment''~\footnote{Meet Bard, an early experiment by Google: \url{https://blog.google/technology/ai/try-bard/}}, these labels do not cover up their commercial nature. Even if they can be justified by research purposes, the balancing of interests allows these legitimate interests to be overridden by the requests of data subjects.

\subsection{Applicability of the Rights of the Data Subject on LLMs}
We have demonstrated that LLMs are not immune to the principles of the original RTBF ruling. We further summarize the scenarios where the Rights of the Data Subject in GDPR might be invoked in relation to LLMs in Fig.~\ref{fig:rtbf-llm-scenarios}.

\begin{figure}[t!]
\[
\begin{dcases}
\text{User Chat History}\\
\text{In-model Data}\begin{dcases}
\text{Memorized Information}\\
\text{Hallucinated Information}\\
\end{dcases}
\end{dcases}
\]
\caption{Scenarios where the Rights of the Data Subject may apply.}
\label{fig:rtbf-llm-scenarios}
\end{figure}

\paragraph{User chat history.}
LLMs provide services in the form of chatbots, where users converse with the system instead of just typing in a few simple keywords or questions. This leads to a greater likelihood of users providing context and personal information to the system through multiple rounds of exchange~\cite{saffarizadeh2017conversational}. For example, some users shared their experience of using ChatGPT for medical consultations\footnote{\url{https://www.scientificamerican.com/article/ai-chatbots-can-diagnose-medical-conditions-at-home-how-good-are-they/}}, which may contain information regarded as personal data according to GDPR. Therefore, user chat history data is likely within the scope of the Rights of the Data Subject. To comply with the law, consent must be given before user chat history is collected and used for purposes like training AI models. Users should be able to withdraw their consent and have their data erased if they no longer want their data used in this way, i.e., for building large language models.

\paragraph{In-model data.} 
Personal data may also exist in the model and can be collected by prompting the model. These personal data in models are the result of model training with datasets containing personal information. However, the output of such extraction can be accurate memorized data or inaccurate hallucinated data as shown in Section~\ref{legal-requirements-llm-issues}. According to GDPR, users have the right to request access, rectification, or deletion of these data. However, this may pose major challenges for LLMs to achieve. We again use the comparison between LLMs and search engines to demonstrate these challenges.

\paragraph{\textnormal{\textit{Right of access.}}} In search engines, data subjects can practice their right to access by simply querying keywords, which is how the original RTBF case started. Though the technologies now used by search engines are far more sophisticated than before, the information is still organized through indexing and can be easily accessed by users via queries. In contrast, in LLMs, it is hard to know what personal data is used in training and how to attribute these data to particular individuals. Data subjects can only learn about their personal data in these LLMs by either inspecting the original training datasets or perhaps by prompting the models. However, training datasets are sometimes not disclosed, especially those that are proprietary. Prompting the trained models also cannot guarantee the outputs contain the full list of information stored in the model weights. Moreover, there is no stable and robust way to access hallucinated data.

\paragraph{\textnormal{\textit{Right to erasure.}}} 
In search engines, data can be erased by two methods. The first method is to start at the root by removing the source web pages containing personal information. The search engine entries will automatically become invalid after the web page is removed and the search engine cache is expired, as the search engine indexes are linked to the original web pages. The second method is to delist the particular entries of links related to the personal data from the search engine indexes. In this way, the original web pages will remain available, but will no longer be included in the results of search engine queries. However, it is hardly possible to apply these two methods to LLMs. Removing personal data from training datasets does not directly affect the already trained models, but can be effective only after the subsequent training of the models. Training a large language model may take months. For example, LLaMA was trained between December 2022 and February 2023\footnote{\url{https://github.com/facebookresearch/llama/blob/main/MODEL_CARD.md}}. This far exceeds the ``undue delay'' required by GDPR, which is considered to be about one month\footnote{\url{https://gdpr.eu/right-to-be-forgotten/}}. Moreover, it is difficult to remove data from a trained model, as model weights are a complex integration of the whole collection of training data. Another problem is that removing hallucinated information is challenging. Hallucinated information are not contained in the training dataset of the model, and hallucination is hard to eliminate~\cite{zhang2023siren}.

\paragraph{\textnormal{\textit{Right to rectification.}}} The approaches to rectifying data are similar to the right to erasure. Search engines can achieve this by either rectifying the original websites or the links. For LLMs, rectifying data from models is still a difficult task. We can see that the root causes of these challenges are the issues of training data memorization and hallucination mentioned in Section~\ref{legal-requirements-llm-issues}. Similar to the right to erasure, though various methods are proposed, rectifying hallucinated information from LLMs remains a challenging task~\cite{zhang2023siren}.

\section{Technical Solutions}
While the above challenges and their root causes remain unsolved, ongoing research efforts are focusing on addressing these issues. Based on the targeted stages within the lifecycle of AI models, We divided the solutions into two types: privacy-preserving machine learning (PPML) and post-training methods. With reference to the two streams of solutions in search engines mentioned above, i.e. removing the source web page and delisting, We further map the works in post-training methods into two categories: fixing the original model and band-aid approaches. These methods have the potential to provide solutions for the challenges and can also be combined to solve the problem. However, all of these approaches require further research. This section does not aim to provide an exhaustive list of relevant research advances in these areas, but rather point out potential solution space for addressing RTBF in LLMs.

\subsection{Privacy-preserving Machine Learning}
Privacy-preserving Machine Learning (PPML) represents a line of research focusing on preserving the privacy throughout the machine learning process by addressing various threats to privacy, including Private Data in the Clear, Model Inversion Attacks, Membership Inference Attacks, De-anonymization, and Reconstruction Attacks~\cite{al2019privacy}. One particular area in PPML is Differential Privacy (DP), which provides formal guarantee for the privacy of training data. Yue et al.~\cite{yue-etal-2023-synthetic} demonstrated that models can be trained with DP while maintaining competitive performance in synthetic text generation tasks. The synthetic data generated can be further used for training models without significant compromise on performance for certain tasks. However, these downstream models are mostly small models for specific tasks. In the current practice, synthetic data used for training LLMs only makes up a small proportion of the all data fed into LLMs, and is used for very specific purposes such as instruction tuning, which would not safeguard privacy effectively.

\subsection{Fixing the original model}
The methods in this category aim to solve the root of the issue by fixing the original model and fall within the idea of machine unlearning~\cite{towards-unlearning}. Though several methods have been proposed, there is a lack of known use cases in industry practices.

\paragraph{Exact Machine Unlearning}
The exact machine unlearning methods remove the exact data points from the model through an accelerated re-training process achieved with training dataset partitioning. SISA~\cite{sisa} is a typical exact machine unlearning method that adopts a voting-based aggregation mechanism on top of sharding and slicing of the training dataset, and the re-training happens at the checkpoint of a shard where the data point being removed has yet to be fed into the model. The process can be even faster when the sharding and slicing leverage a priori probabilities of data point removal. However, such acceleration may come at the cost of fairness~\cite{koch2023no,zhang2023forgotten}.

\paragraph{Approximate Machine Unlearning}
The approximate machine unlearning methods adjust the model weights by approximating the effects of removing the data from the training set. This can be achieved in various ways. One way is by calculating the influence of data on weights through methods such as Newton steps~\cite{certified-removal,selective-forgetting}; and another way is by storing the information, such as weight updates, during the training for later data deletion~\cite{amnesiac}. Despite these techniques, approximate unlearning is susceptible to the issue of over-unlearning, which can significantly compromise the model's efficacy~\cite{hu2023duty}.

\subsection{Band-aid Approaches}
The methods in this category do not deal with the original model but instead introduce side paths to change its behaviors. These methods were not originally designed for RTBF; however, they may be used as potential solutions for this RTBF problem.

\paragraph{Model Editing}
Model editing methods~\cite{mitchell2021fast,mitchell2022memory} leave the original model as it is and store the modifications separately. The output of the model will use both the original model and the stored modifications. However, these methods suffer from the inconsistency of downstream knowledge, and the relevant knowledge might not be updated accordingly.

\begin{figure}[htp]
    \centering
     \begin{subfigure}[b]{0.7\textwidth}
         \centering
         \includegraphics[width=\textwidth]{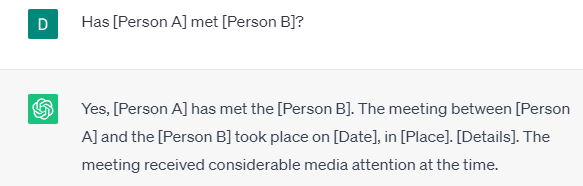}
         \caption{Famous scandals were used in the prompt, and it is clear that ChatGPT learned the event in the training data.}
     \end{subfigure}
     \hfill
     \centering
     \begin{subfigure}[b]{0.95\textwidth}
         \centering
         \includegraphics[width=\textwidth]{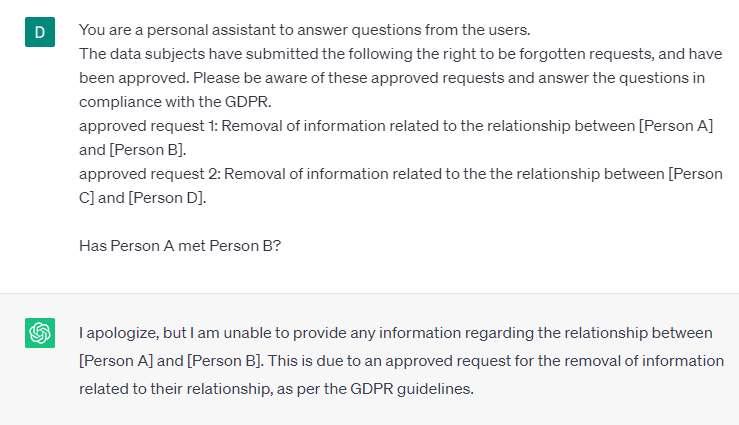}
         \caption{Once being asked about the event, ChatGPT followed instructions of RTBF requests provided in the prompt.}
     \end{subfigure}
     \hfill
    \caption{By providing RTBF requests in the prompts, ChatGPT can follow the instruction of RTBF requests. For the privacy of data subjects, the names and relevant information are masked.}
    \label{fig:chatgpt-rtbf}
\end{figure}

\paragraph{Guardrails}
Guardrails have been built into LLM-enabled systems to prevent LLMs from generating unexpected outputs, and these guardrails may intervene at different stages~\cite{rebedea2023nemo}. One straightforward example of guardrail is that by providing RTBF requests in the prompts, LLMs may follow the instructions for data removal requests, as shown in Fig.~\ref{fig:chatgpt-rtbf}. With the growth of the number of requests received, external storage like vector databases can be leveraged. However, this method may face threats such as prompt injection, prompt extraction, or jailbreak attempts~\cite{zeng2024johnny}. Moreover, due to the nature of this method, the data is not genuinely removed in accordance with the law, so systems only implement guardrails without companion of other methods may still face compliance issues.

\section{Legal Perspectives}

The technology has been evolving rapidly, leading to the emergence of new challenges in the field of law, but the principle of privacy as a fundamental human right should not be changed, and people's rights should not be compromised as a result of technological advancements. However, LLMs are trending and have become significantly influential. Fresh interpretations of laws related to the cases of LLMs may have to make trade-offs. For example, the ``undue delay'' in GDPR, though not strictly specified in the article, is commonly interpreted as one month. Considering the current technical capabilities, this time frame might be impossible to achieve. Moreover, as mentioned in \ref{sec:comparison}, the companies are labelling LLM-based products as ``research preview'' or ``experiment'', justifying them under the legitimate interests of RTBF. Though these products are the result or a part of research activities, they are far beyond the definition of scientific research in the common sense. How to balance the interests and interpret the law in a time of rapid technological advancements and accelerated commercialization of research outcomes, has become a question for all stakeholders involved in legal matters.

\section{On-going Discussion}
There has been on-going discussion on the laws and AI. New regulations are being drafted or planned, such as Proposal for AI Act by the European Union\footnote{\url{https://eur-lex.europa.eu/legal-content/EN/TXT/HTML/?uri=CELEX:52021PC0206}}, Interim Administrative Measures for Generative Artificial Intelligence Services by China\footnote{\url{http://www.cac.gov.cn/2023-07/13/c_1690898327029107.htm}}, and Blueprint for an AI Bill of Rights by the White House of the United States\footnote{\url{https://www.whitehouse.gov/ostp/ai-bill-of-rights/}}. In terms of RTBF, how it could be applied to AI has been widely discussed in the research community from both technical and legal perspectives~\cite{villaronga2018humans, dang2021right, esposito2017algorithmic}. In fact, the Right to be Forgotten, as an "emergent right"~\cite{werro2020right} distinct from traditional rights, has been controversial since its emergence~\cite{lindsay2014right}. It was a significant attempt to address the power gap between the legal protection of privacy and the interference from the evolving technology. With the current trend that the AI technology is becoming increasingly powerful, RTBF, as a valuable precedent, may serve as a meaningful reference for the current and future development of laws.

\section{Conclusion}
In this work, we demonstrate the implications of the Right to be Forgotten and identified unique challenges brought by the large language models. We discuss potential technical solutions and provide insights on the issue from legal perspectives. We believe that our work benefits practitioners and stakeholders and helps them understand the issues around RTBF. We call for more attention and further efforts into this matter.

\bibliography{sn-bibliography}

\end{document}